\documentclass[11pt,preprint]{aastex}
\usepackage{graphicx}
\usepackage{natbib}
\usepackage{amsmath}
\usepackage{multirow}
\usepackage{array}
\usepackage{subfigure}
\usepackage{morefloats}
\usepackage{xcolor}
\newcommand{\be}{\begin{equation}}
\newcommand{\ee}{\end{equation}}

\newcommand{\energy}{{\cal E}} 
\newcommand{\uvec}{{ {\bf u} }}

\newcommand{\sigmauv}{{ \sigma_{\rm uv} }}
\def\lta{\,\raise 0.3 ex\hbox{$ < $}\kern -0.75 em
	\lower 0.7 ex\hbox{$\sim$}\,}
\def\gta{\,\raise 0.3 ex\hbox{$ > $}\kern -0.75 em
	\lower 0.7 ex\hbox{$\sim$}\,} 

\begin{document}

\title{Survival of Primordial Planetary Atmospheres: Mass Loss from Temperate Terrestrial Planets}

\author{Alex R. Howe$^1$}
\author{Fred C. Adams$^{2,3}$}
\author{Michael R. Meyer$^{2}$}
\affil{$^{1}$NASA Goddard Space Flight Center, 8800 Greenbelt Rd, Greenbelt, MD 20771, USA}
\affil{$^{2}$Department of Astronomy, University of Michigan, Ann Arbor,
	Michigan 48109, USA}
\affil{$^{3}$Department of Physics, University of Michigan, Ann Arbor,
Michigan 48109, USA}


\begin{abstract}

The most widely-studied mechanism of mass loss from extrasolar planets is photoevaporation via XUV ionization, primarily in the context of highly irradiated planets. However, the EUV dissociation of hydrogen molecules can also theoretically drive atmospheric evaporation on low-mass planets. For temperate planets such as the early Earth, impact erosion is expected to dominate in the traditional planetesimal accretion model, but it would be greatly reduced in pebble accretion scenarios, allowing other mass loss processes to be major contributors. We apply the same prescription for photoionization to this photodissociation mechanism and compare it to an analysis of other possible sources of mass loss in pebble accretion scenarios. We find that there is not a clear path to evaporating the primordial atmosphere accreted by an early Earth analog in a pebble accretion scenario. Impact erosion could remove $\sim$2,300 bars of hydrogen if 1\% of the planet's mass is accreted as planetesimals, while the combined photoevaporation processes could evaporate $\sim$750 bars of hydrogen. Photodissociation is likely a subdominant, but significant component of mass loss. Similar results apply to super-Earths and mini-Neptunes. This mechanism could also preferentially remove hydrogen from a planet's primordial atmosphere, thereby leaving a larger abundance of primordial water compared to standard dry formation models. We discuss the implications of these results for models of rocky planet formation including Earth's formation and the possible application of this analysis to mass loss from observed exoplanets.

\end{abstract}

\maketitle


\section{Introduction}

Conventional planet formation models assume the build-up of isolation
mass cores from the growth of solids into planetesimals (e.g.,
\citealt{Safronov69,Wetherill93,Goldreich04,Benz14}). In the classical
picture, isolation mass oligarchs (with masses comparable to Mars and
semimajor axes $a\sim1$ AU) are built up through collisions of
planetesimals aided by gravitational focusing. They collide violently
in the chaotic growth phase after dissipative gas is largely gone at
system ages 10 -- 100 Myr. These planets may form without any
primordial atmospheres if the isolation mass is reached after the gas
disk dissipates. If they do accrete atmospheres, however,
their subsequent thermal evolution and mass loss are almost certainly
dominated by giant impacts \citep{BS18}. Building up an Earth-mass
planet through successive giant impacts would likely have removed any
primordial volatiles, and any thin, primordial atmosphere surviving
these impacts could be quickly lost to photoevaporation
\citep{Johnstone19}.

An important alternative theory of planet formation involves the
streaming instability and pebble accretion, which can form terrestrial
planets faster than planetesimal accretion, i.e., while the
circumstellar disk is still gas rich (e.g., \citealt{Bitsch15,Chambers18}). 
In a pebble-accretion scenario, a super-Earth-mass planet could form
within 1 Myr and capture a much deeper gas-rich envelope. Estimates
from \cite{Ginzburg16} suggest that mass fractions up to 2\% could be
realized in a hydrogen- and helium-rich atmosphere. If the nebula
is ``wet'' (e.g., \citealt{Ciesla06}) due to the migration of small icy
bodies interior to the ice-line (and subsequent sublimation of
volatiles into the gas phase), an Earth-mass planet could capture a
water mass fraction approaching $2\times10^{-4}$, comparable to the
present-day Earth within a factor of 3 \citep{Meech15}.

It is not clear that the above results are consistent with formation
models explaining the inner solar system planets, as discussed
below. Nonetheless, it is of interest whether there are mechanisms
that could deplete light elements in the primordial atmosphere. In
this paper, we explore processes that could contribute to mass loss
over the lifetime of such a
planet. Most notably, we explore the possibility that molecular
photodissociation by ultraviolet light could be an important source of
mass loss. This mechanism has not been explored thoroughly in the
context of mass loss from young exoplanets. This process should be
effective, however, because photodissociation of hydrogen occurs at lower energies than photoionization \citep{Draine96,Heays17}, and thus would increase the ultraviolet flux available for upper atmosphere heating and escape. \cite{Draine96} also estimate the broad-spectrum efficiency of photodissociation at $\sim$15\%, which would also increase the total mass loss over photoionization alone. In contrast, mass
loss on exoplanets is usually modeled based only on photoionization of
hydrogen atoms \citep{Watson81}. For completeness, we also investigate other sources of mass loss and
quantify their relative importance.

A related question is whether this mass loss of hydrogen and helium
will leave behind any of the water accreted from the disk. It is an
interesting coincidence that the expected amount of water accreted
from the disk is similar to the mass of Earth's oceans. This finding
may indeed be mere coincidence because for many years, isotopic
evidence has indicated that the majority of the water in Earth's
oceans must have originated from beyond the ice line. In particular,
the D/H ratio is indicative of cold cloud chemistry, such as that
observed in molecular cloud cores (and presumably delivered to the
outer nebula during the early phases of star formation). On the other
hand, measurements of D/H ratios in deep mantle lava are closer to
the solar ratio, indicating the possibility of a deep mantle reservoir
of primordial water \citep{VD14,Hallis15}, which could have been
accreted from the primordial circumstellar disk. While this
possibility remains speculative, exploring mass loss processes that
could leave primordial water behind could shed light on whether these
findings are consistent with pebble accretion formation models.

The pebble accretion scenario has additional complications. Notably,
pebble accretion has a natural endpoint at a super-Earth mass (e.g. \citealt{Bitsch15,Johensen17}). In
order to produce an Earth-mass planet, the accumulation phase must be
interrupted by some other mechanism, and this process may also affect
the volatile content of the planet. Also, pebble accretion cannot be
the only process involved in the formation of our solar system's
terrestrial planets because any theory of planet formation must still
explain the Moon-forming impact and Mercury's iron-rich composition (e.g. \citealt{Ward00,Benz07}).
Nonetheless, we can place reasonable limits on the extent of these
other processes, and we find that they may be comparable to photoevaporation for planets with relatively low insolation like early Earth, but they probably do not dominate mass loss in
most cases.

This paper is structured as follows. In Section \ref{literature}, we
summarize the current literature on mass loss from exoplanets,
particularly to justify our addition of photodissociation to the list
of viable processes.  In Section \ref{massloss}, we compute the
anticipated mass loss from several mechanisms, including both
photoionization and photodissociation, along with a summary of the
published results for impact erosion and their potential application
to a pebble accretion scenario. In Section \ref{discussion}, we apply
these results to observable properties of exoplanets and discuss their
implications for formation models of planets and planetary systems. We
summarize our findings in Section \ref{summary}.

\section{Summary of Literature}
\label{literature}

Most studies of atmospheric loss from extrasolar planets are based
directly or indirectly on the work of \cite{Watson81}. This work
considered energy-limited mass loss due to heating by extreme
ultraviolet (EUV) radiation from the star. This mechanism is dominant
for planets that are highly irradiated, including a large fraction of
known exoplanets. Although this paper considers additional mass loss
mechanisms such as Jeans escape and thermal winds, we focus primarily
on photoevaporation, which continues to dominate even at Earth's level
of insolation when impact erosion is not considered.

\cite{Watson81} did not define the term ``EUV''. Although they
describe the limit of efficient EUV heating to take place when the gas
is ionized, they describe the heating only in terms of absorption,
rather than photoexitation. EUV radiation is conventionally
taken to be the radiation blueward of either 121 nm or 91.2 nm, meaning that this convention does not necessarily faithfully model the ionizing flux, which is strictly blueward of 91.2 nm. Dissociation of molecular hydrogen occurs primarily via line processes blueward of 111 nm \citep{Draine96}, so dissociating radiation may or may not be included in the definition of ``EUV'' depending on the context.

In addition, most studies of atmospheric loss from exoplanets use
directly or indirectly the estimates of \cite{Ribas05} for the XUV
flux as a function of time for solar-analog stars. This model
integrated the X-ray and UV flux over the wavelength range 0.1 -- 120
nm, again not modeling the exact ionizing flux. This previous work
also did not include stars of other spectral types. However, some studies used different models of XUV flux, e.g. \citealt{Murray-Clay09,Lammer14}.
Meanwhile, photodissociation has been addressed for exoplanets in other contexts such as water loss \citep{Jura04}, but very little in the context of hydrogen-rich atmosphere loss.

While \cite{Watson81} dealt specifically with Earth and Venus, the
majority of scholarship on atmospheric loss from exoplanets has
focused on hot jupiters (e.g.,
\citealt{Lammer03,Baraffe04,Hubbard07}). (Here, we are taking ``hot jupiters'' generically to be planets with equilibrium temperatures $>$1,000 K and mass greater than Saturn, $\gtrsim0.3$ M$_J$.) Nonetheless, studies of mass loss from hot
jupiters often do not clearly specify whether photoionization or
photodissociation processes are considered.



\cite{Murray-Clay09} modeled mass loss on hot jupiters incorporating a
number of non-thermal processes. They did consider dissociation, but
they only considered thermal dissociation, not photodissociation, and
they concluded that the temperatures involved were high enough for the hydrogen to be fully thermally dissociated. For their specific model of
energy-limited ultraviolet heating, they explicitly considered only
photoionization, and as such included only UV radiation blueward of
91.2 nm.


\cite{Lopez12} and their subsequent papers studied mass loss from super-Earths and
mini-Neptunes. However, they also cited the \cite{Ribas05} model for XUV flux, and they
explicitly described it as modeling only ionization, not
dissociation. Following from this model, \cite{Jin14} described the
problem in the same manner while recreating the models of
\cite{Baraffe04} and \cite{Lopez12}. Meanwhile, \cite{Howe15} did
mention both ionization and dissociation as possible pathways of XUV
heating, but they also used the \cite{Ribas05}.

Finally, \cite{Lammer14} considered both ionization and dissociation
processes. (They also made some mention of ``FUV'' fluxes, although
they did not define the term.) However, their mass loss model was based
only on XUV fluxes and thus also did not take photodissociation into
account in practice.


The above discussion thus indicates that photodissociation-induced
atmospheric evaporation has received little attention
in applications to exoplanets. Nonetheless, the energy levels involved
indicate that this process should be taken into account when modeling
exoplanet evolution. In this case, to leading order, the same model
for mass loss applies, except that the wavelength range modeled should be set to cover the full range of dissociating radiation in addition to ionizing radiation, and the efficiency factor should
be adjusted accordingly.


\section{Estimates of Mass Loss}
\label{massloss}


For this analysis of mass loss, we consider pebble accretion taking
place within the ice line \citep{Chambers16} of a circumstellar disk.
This scenario results in the rapid formation of a planet, which is
assumed to have the same mass and orbital distance as Earth. As an
initial condition, the planet captures an additional 2\% of its mass
from the gas disk before dissipation \citep{Ginzburg16}. The resulting
0.02 $M_\earth$ atmosphere corresponds to a surface pressure of
$\sim$23,000 bar, which we use to quantify the mass loss from the
various processes. (For comparison, one ocean in the form of water
vapor would be $\sim$230 bar.) Such an atmosphere will be optically
thick, keeping the surface hot with an estimated surface temperature
of $\sim$4,500 K \citep{Popovas18}, and with a Kelvin-Helmholtz cooling
time of $\sim$3 Myr \citep{MR09-Collisions}. These parameters represent our starting
conditions for purely thermodynamic loss processes such as Jeans
escape and thermal winds.

At this temperature, the atmosphere is
likely to be inflated to a radius $\sim2\,R_\oplus$. As an order of magnitude estimate, the scale height of a hydrogen atmosphere will be $\sim$200 km, and the height of the atmosphere will be 30-40 scale heights from the surface to the exobase, resulting in a total height of $\sim1\,R_\oplus$. The actual height of the atmosphere must be determined by a numerical calculation, which is beyond the scope of this paper (but cf. \citet{Howe14} at high entropy levels). As a result, we assume an initial exosphere radius of $R_{exo}=12,000$ km in our model. For
comparison, planetesimal accretion models predict the planet to be
repeatedly heated to a temperature $\sim$1,500 K (about 6-9 times)
by giant impacts during its formation phase, thereby resulting in a 
thinner atmosphere and a shorter cooling time.


A number of different processes can contribute to atmospheric mass
loss, although not all authors agree on the nomenclature (e.g. \citealt{Catling09,Catling17}). Here, we group the
relevant mass loss mechanisms into two broad categories, including 
(i) thermal escape and (ii) atmospheric erosion. The first class of
mechanisms occurs when the atmosphere is heated, causing the
constituent molecules to escape into space. Thermal escape can be
considered in the limit where individual molecules escape from a
collisionless exosphere (Jeans escape, Section 3.1), or when the
outflow takes place in the fluid limit and is driven by atmospheric
heating (hydrodynamic escape, Section 3.2). In this latter case,
different heating mechanisms come into play, where this paper
considers approximate treatments of both ionization heating (Section
3.5) and dissociation heating (Section 3.6). The second class of mechanisms, atmospheric erosion,
includes ablation by stellar winds (Section 3.3) and impact erosion,
which takes place as large bodies impinge upon the atmosphere (Section
3.4).\footnote{For completeness, we note that supra-thermal escape can
also occur. In this setting, individual particles are accelerated to
escape velocity due to chemical reactions or ionic interactions. These
mechanisms are generally subdominant and are not considered here
\citep{Catling17}.}  These mass loss mechanisms are depicted
schematically in Figure 1.


\begin{figure*}[htbp]
	\includegraphics[width=0.99\textwidth]{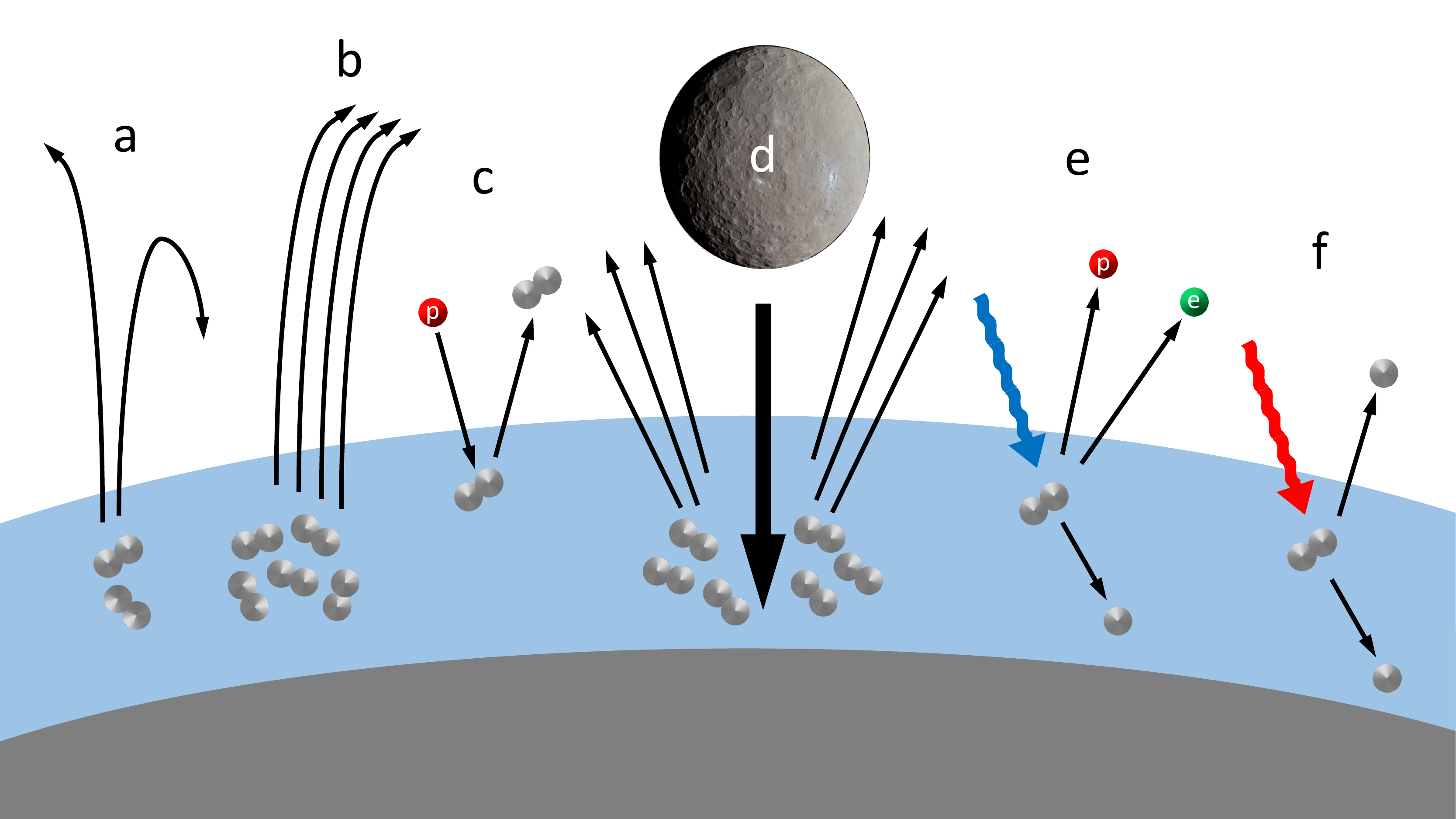}
	\caption{Diagram of the molecular processes driving the mass loss mechanisms we investigate in this paper: (a) Jeans escape, (b) hydrodynamic winds, (c) stellar wind ablation, (d) impact erosion, (e) photoionization, and (f) photodissociation. The colors of light are arbitrary and have been chosen to illustrate the difference in wavelength between photoionization and photodissociation. Note that photoevaporation processes occur in the collisional region of the atmospheres and therefor involve heating of the atmosphere rather than ejection of individual particles by photons.}
	\label{processes}
\end{figure*}

\subsection{Jeans Escape}
\label{jeans}

The first means of atmospheric mass-loss is Jeans escape. In this
case, the atoms and/or molecules in the high-velocity tail of the
Maxwell-Boltzmann distribution are moving fast enough to escape the
gravity well of the planet \citep{Catling09}. The particles in this high-energy tail
continually escape from the top of the atmosphere. Jeans escape is
considered significant if the escape velocity is less than about six
times the mean molecular speed, which occurs on Earth only for
hydrogen and helium.


Jeans escape depends on the temperature of the exosphere and is thus
exponentially increased when the young planet is heated by rapid
accretion. The rate of Jeans escape (in molecules per unit area per
unit time) can be derived from the Maxwell-Boltzmann distribution and
is given by  
\begin{equation}
\Phi_J = \frac{1}{2\sqrt{\pi}}nv_0(1+\lambda_J)e^{-\lambda_J},
\end{equation}
where $v_0 = (2kT/m)^{1/2}$ is the most probable molecular speed, and
$\lambda_J$ is the escape parameter, given by $\lambda_J \equiv$ 
$GMm/(kTR_{exo})$. The density of the exobase, $n$, is determined by
setting the scale height equal to the mean free path, i.e., 
\begin{eqnarray}
\frac{kT}{mg} = \frac{1}{\sqrt{2}\sigma n}, \\
n = \frac{mg}{\sqrt{2}kT\sigma},
\end{eqnarray}
where $\sigma=\pi d^2/4$ is the cross section of the molecules, and $m$ is the
mean molecular weight. At a temperature of $T$ = 4,500 K, the hydrogen
gas will be dissociated (which will not be the case at later times when photoexcitation becomes more important), so we must use the values for atomic
hydrogen with a molecular weight of $m=1$ u and an atomic diameter of $d = 106$ picometers.


Because of the low density of molecules at the exobase, Jeans escape
is slow, and relatively little mass can be removed from the thick
atmosphere in our model.  Even at very high temperatures, the rate of escape increase only slowly with temperature, scaling with $T^{1/2}$,
so this conclusion would remain valid even if our temperature estimate
is too low. In this model, if we assume $R_{exo} = 12,000$ km, then
the mass loss rate of hydrogen from the planet is $3.4\times10^7$ g/s.
For a time span of 3 Myr, the total mass loss would be about
$5.4\times10^{-7} M_\oplus$. In terms of pressure, this result implies
that Jeans escape is sufficient to evaporate $\sim$0.6 bars of hydrogen
from an Earth-like primordial atmosphere.


\subsection{General Hydrodynamic Escape}

In the limit where the gas is coupled (the fluid limit), 
hydrodynamic winds provide an important description of mass 
loss. These winds can remove mass
from the upper atmosphere at a rate determined by the temperature of
the atmosphere. We can estimate the outflow rate for the case of an
isothermal atmosphere, which is expected to be a good approximation to
a planet's upper atmosphere, based on the requirement that the flow
must pass smoothly through the sonic point\footnote{Whereas this section presents the general mechanism
for hydrodynamical escape, later sections consider specific 
heating mechanisms, but utilize an approximate description of 
the dynamics (Sections 3.5, 3.6).} Other forces can further suppress hydrodynamic escape such as magnetic fields \citep{OwenAdams}, but these are beyond the scope of this paper. The full derivation of
these results is given in Appendix \ref{derivation}. In the isothermal
approximation, the thermal wind from the atmosphere is given by 
\be 
{\dot M} = 4\pi r^2 \rho v = 4\pi R_b^2 \rho_b a_s \lambda_{th} \,,
\label{hydroMdot}
\ee
where $\rho_b$ is the density at the base of the outflow, $R_b$, $a_s$ is the sound speed,
and the parameter $\lambda_{th}$ is a function of the dimensionless
potential $b$. To first order, this parameterization can be
approximated with the expressions 
\begin{align}
b &= {G M_P \over R_b a_s^2} \approx 62.5 
\left(\frac{R_b}{R_\oplus}\right)^{-1}\left( {a_s \over 1\, {\rm km} \,\,{\rm s}^{-1}} \right)^{-2} \, \\
\lambda_{th} &\approx { {\rm e}^{3/2} \over 4} b^2 {\rm e}^{-b}\,. 
\end{align}

To evaluate the mass loss rate from Equation (\ref{hydroMdot}), we 
need to determine the number density $n_b$ at the base of 
the flow. For example, when the mass loss is driven by 
incoming UV photons, the base occurs where the incoming 
radiation becomes optically thick and the expression for 
mass loss becomes
\be
{\dot M} = (5.3\times10^9 {\rm g}\,\, {\rm s}^{-1}) \left(\frac{R_b}{R_\oplus}\right)^{1/2} b^{5/2} {\rm e}^{-b}\,,
\label{outflow} 
\ee
where the values use to compute this are specified in the Appendix. (Note that this assumes $R_b = R_\oplus$.)

Because the parameter $b$ depends on the sound speed, it is
temperature-dependent. Specifically, $b\propto (R_PT)^{-1}$. For typical exospheric properties of $R_b = R_\oplus$ and $T=1500$ K predicted by planetesimal accretion models, $a_s\approx2.5$ km s$^{-1}$, and $b\approx10$. If we define the prefactor $C = b^{5/2}e^{-b}$, then for $b=10$, $C \approx 0.014$. $C$ takes on a maximum value of $\approx$0.81 for $b\approx 2.5$. At higher temperatures, the flow does not transition smoothly through the sonic point and will be time-dependent, or will
take on the form of a shock, while at low temperatures, it is exponentially suppressed.

In this paper, we consider two major regimes of hydrodynamic escape. First, there is the short-lived, large-radius, high-temperature state caused by heating due to pebble accretion or giant impacts. For conditions of $T$ = 3,000 K and $R_b = 2\,R_\oplus$, $C$ approaches its maximum value of 0.8, which yields a mass loss over 3 Myr of $1.0\times10^{-4}\,M_\oplus$, corresponding to 118 bars of hydrogen removed from an Earth-like primordial atmosphere.

Second, we consider a quasi-steady state outflow due to ultraviolet heating over the lifetime of the planet. This regime is extremely temperature-sensitive, but if we assume optimistic values for this regime of $T$ = 1,500 K and $R_b = 1.5\, R_\oplus$, then $C = 0.15$, then the integrated mass loss over 5 Gyr is 0.026 $M_\oplus$, which is greater than the initial hydrogen content of our model and justifies using the energy-limited approximation for computing mass loss up to the order of this quantity.



\subsection{Ablation by Stellar Winds}

Stellar winds can erode planetary atmospheres by directly imparting
momentum to the upper atmosphere through the action of wind particles.
This effect can be calculated from the mass loss rate and speed of the
stellar wind. The mass loss rate from young stars as a function of
time can be described with a simple model of the form \citep{Skumanich81}, 
\be
{\dot M}_\ast = {\dot M}_0 \left( {t_0 \over t_0 + t} \right)^2\,,
\ee
where the benchmark mass loss rate for Sun-like stars is ${\dot M}_0
\approx 2\times10^{-11}\,M_\odot$ yr$^{-1}$, and the time scale $t_0$
= 100 Myr \citep{Wood02}. We can find an upper limit to the planetary
mass loss driven (directly) by the stellar wind by first finding the
total amount of (stellar) mass that flows through the volume subtended
by the planet, i.e., 
\be
\Delta M = \int_0^\infty dt {{\dot M}_0 \over 4\pi d^2}\pi R_p^2 \left( {t_0 \over t_0+t} \right)^2 = {{\dot M}_0 t_0 R_p^2 \over 4 d^2} \,,
\ee
where $R_p$ is the radius of the planet, and $d$ is the radius of the
planetary orbit (assumed here to be circular). The total mass 
$\Delta M$ that intercepts the planet is thus about 
$\Delta M\sim2\times10^{21}$ g. Notice that we are assuming 
$R_p\sim1$ R$_\oplus$ instead of the exospheric radius because we are
considering the problem over a longer time scale of 100 Myr, over
which time the planet would cool, and the atmosphere would compress so that the effective radius will be near 
that of the solid surface. 

The amount of mass that can be directly removed by the incoming wind
is limited by conservation of momentum: $\Delta M_p = \frac{v_w}{v_e}\Delta M_*$. Since the wind speed is about
$v_w\sim300$ km/s, and the escape speed from the planet is about
$v_e\sim11$ km s$^{-1}$, the incoming mass $\Delta M$ could (at most)
remove a mass of
$\sim5\times10^{22}\,\rm{g}\sim8\times10^{-6}\,M_\oplus$. Stellar wind
ablation can remove only 10 bars of hydrogen from an Earth-like
primordial atmosphere, much less than any other processes under 
consideration. 


\subsection{Impact Erosion}
\label{impact}

While giant impacts are generally a feature of planetesimal formation
models, it is clear that they will occur regardless of the formation
mechanism. In particular, the mass loss caused by giant impacts in the
context of the Moon-forming collision has been studied for some time,
with a range of results. For example, \cite{Ahrens93}, implied that
the Moon-forming impact could have unbound Earth's entire atmosphere
by itself, whereas \cite{Genda03} estimated only a 20\% atmosphere
loss. However, the general theory of mass loss caused by possible
multiple giant impacts of a range of sizes is more complex.

Giant impacts can cause mass loss from planets in two ways: [1] The
direct mechanical ejection of a large fraction of the atmosphere, and
[2] Through the thermal wind induced by the heating of the remaining
atmosphere. Both of these mechanisms can be major contributors to mass
loss, with magnitudes comparable to or greater than the other
mechanisms under consideration.

The mechanical ejection of the atmosphere by impacts over a wide range
of impactor sizes was studied by \cite{Schlichting15}. They identified
four regimes based on the size of the impactor and additional energy
considerations. For small impactors, the shock generated by the impact
(or airburst) is not strong enough to eject any of the atmosphere.
Slightly larger impacts may be approximated such that they can remove
all of the atmosphere up to a certain airmass, thus ejecting a cone
above the impact point. The threshold for this ejection is approximated such that the mass of solid ejecta ($M_{ej}\sim M_{imp}$) is greater than the atmosphere mass per unit solid angle. As such, the minimum impactor size for mass
ejection varies with the thickness of the atmosphere. For an
Earth-like atmosphere, quite small impacts can reach this threshold,
but for a primordial atmosphere with a mass of 0.02 $M_\earth$, this
minimum size is $r\approx25$ km.

As the impactor size grows, the ejected cone widens until it reaches
the horizon, thus ejecting the entire spherical cap above the tangent
plane to the impact. For our primordial atmosphere model, this
critical impactor size is $r\approx600$ km. In the third regime, from
this size scale up to $r\approx1,000$ km, the amount of mass loss is
constant as a function of impactor size, and is still determined by
the spherical cap above the tangent plane. Finally, for truly giant
impactors of $r\gtrsim1,000$ km, regardless of atmosphere mass, direct
transfer of momentum through the solid mass of the planet will eject
significantly more of the atmosphere than the spherical cap \citep{Schlichting15}.

Planetesimal accretion predicts Mars-sized impactors as a distinct
population. Interestingly, these giant impacts are not predicted to
eject the entire atmosphere. Indeed, even with equal-mass impactors,
tens of percent of their atmospheres would be ejected, but not the
entire atmosphere \citep{Schlichting15}. Ten Mars-sized impactors in sequence, however,
could plausibly remove virtually all of the primordial atmosphere from
an Earth-mass planet over the course of planet formation. In this
case, any remaining atmosphere would have to be produced via
outgassing.

Another surprising result of \cite{Schlichting15} is that while total
mass lost will be dominated by giant impacts, smaller bodies produce
the most efficient atmosphere stripping in terms of mass of gas
ejected per unit mass of impactor. The optimal case occurs for small
impactors near the minimum size for ejection, which have an ejection
efficiency of $\sim$20\%. This extremal case provides an upper bound
for impact erosion that can be also be applied to pebble accretion: no
more than 20\% of the mass accreted in the form of planetesimals (with $r>25$ km in our example) will be ejected from the atmosphere.



The other important process in impact erosion is the thermal wind
induced by the heating of the atmosphere after the impact
\citep{BS18}. This heating would also inflate the atmosphere, not just
by several Earth radii, but potentially all the way to the Bondi
radius at tens of Earth radii (or the Hill radius for close-in
planets, for which it is smaller). Such an extended atmosphere will thermally 
evaporate much more quickly than any of the processes we study in this work. 
However, the amount of mass lost via thermal wind following a giant impact depends on the
base temperature of the atmosphere, for which models indicate a wide
range of possibilities. \cite{BS18} modeled scenarios with base temperatures ranging
from 2,000 K, for which the mass loss is negligible, to 10,000 K, for
which almost the entire envelope is evaporated in $\sim$2 Myr (with a
comparable cooling timescale). While these numbers are approximate,
for the most likely temperatures it appears that mechanical ejection
is dominant over thermal winds.

Unfortunately, the left-over mass from pebble accretion that would
form into planetesimals is not well-studied, and its quantity is
uncertain. Planetesimal accretion models often postulate a ``late veneer'' scenario (e.g. \citealt{Schlichting12}), in which Earth accreted
an additional $\sim$1\% of its mass from small bodies after the final
assembly of the planets. However, this is not necessarily a good guide
for the pebble accretion scenario we consider because models that invoke pebble accretion can form planets much faster, in less than 1 Myr, and their solid particle dynamics are very different, being influenced by the gas disk, given that typical gas disk lifetimes are 3 Myr for sun-like stars. In this model, we consider only impacts occurring after the dissipation of the gas disk, i.e. after the atmosphere has finished accreting.

Planetesimal accretion and pebble accretion can potentially coexist in comparable amounts during the gas disk lifetime \citep{Schoonenberg19}, but {\it late-stage} planetesimal accretion in a pebble accretion model is expected to be $\lesssim1\%$ (cf. \citealt{Madhu17} and Fig. 7 of \citealt{Liu19}). Thus, we adopt a value of 0.01 $M_\earth$ of planetesimals
accreted within our pebble accretion scenario as a plausible upper bound, with
the caveat that the true number could differ by an order of
magnitude. If this mass increment in planetesimals is deposited on the
planetary surface with an ejection efficiency of 20\%, including
thermal winds, 0.002 $M_\earth$ of gas will be lost. In this toy
model, impact erosion will remove 2,300 bars of hydrogen from the
planet, any Moon-forming impacts notwithstanding.


\subsection{XUV Photo-Ionization and Evaporation}
\label{ionize}

The standard prescription for mass loss on super-Earths, due to \citealt{Watson81}, is to assume an energy-limited approximation for ionization by
XUV photons, using a specific efficiency factor, usually $\sim$10\%. This is an optimistic approximation for hydrodynamic escape, but it often applies for low stellar fluxes (e.g. \citealt{Murray-Clay09,OwenWu17}), so we likewise use is as an optimistic approximation for our analysis.

To compute the mass loss rate due to photoevaporation, consider that the escape
energy per unit mass is given by the gravitational potential, $E_{esc}
= GM_p/R$, and the energy intercepted by the planet is 
$\pi F_{XUV}R_{XUV}^2$. Assuming an efficiency factor $\epsilon$, 
the total mass loss rate is then given by 
\begin{equation}
\label{eq:ion}
\dot{M} = \frac{\epsilon\pi F_{XUV}R_{XUV}^3}{GM_p},
\end{equation}
where $R_{XUV}$ is the radius of the planet at the altitude at which
XUV photons are absorbed. For a hydrogen atmosphere in Earth gravity, the UV cross section for atomic hydrogen is $\sigma_{uv} = 2\times10^{-18}$ cm$^2$, and the photosphere occurs at a pressure of $P = 0.67/\sigma_{uv}\times\mu g = 1.1$ nbar. Note that this assumes a certain percentage of
the intercepted XUV energy is converted to kinetic energy of lost
particles (and consequently encapsulated by the efficiency factor), 
not just one atom lost per XUV photon. A tidal correction must be
applied to planets orbiting very close to their parent stars, but for
our model, we assume it to be negligible.

Because we wish to express the mass loss in terms of surface pressure,
this expression can be simplified. The mass of the planet cancels
out, so that for a cumulative XUV flux, $F_{cum}$, impinging on the 
atmosphere, the integrated atmospheric loss is 
\begin{eqnarray}
\Delta P = \Delta M\frac{g}{4\pi R_p^2} = -\frac{\epsilon F_{cum}R_{XUV}^3}{4R_p^4} \approx -\frac{\epsilon F_{cum}}{4R_p}& \\
               = -(39\,{\rm bar})\,\left(\frac{\epsilon}{0.1}\right)\left(\frac{F_{cum}}{10^{18}\,{\rm erg\,cm^{-2}}}\right)&\left(\frac{R_{XUV}}{ R_\oplus}\right)^3\left(\frac{R_\oplus}{R_p}\right)^4, \nonumber
\end{eqnarray}
where ``1 bar'' represents an atmosphere mass of $\frac{4\pi R_\oplus^4}{GM_\oplus\times1\,bar}$.

The XUV flux from a Solar-type star is estimated at 1 AU by Ribas et al. (2005) as a function of age in Gyr, $t_9$:
\begin{eqnarray}
F_{XUV} = 504\,\rm{erg\,s^{-1} cm^{-2}}, t < 100 \rm{Myr} \\
F_{XUV} = 29.7 t_9^{-1.23}\,\rm{erg\,s^{-1} cm^{-2}}, t > 100 \rm{Myr}.
\label{eq:ionloss}
\end{eqnarray}
Note that this expression is an overestimate because it covers the
wavelength range of 1 -- 118 nm rather than the 1 -- 91 nm of interest
here (although this range will become relevant again in Section
\ref{dissociate}). However, we use this estimate here because it is
the standard for modeling mass loss from irradiated exoplanets. For our model, we assume R$_{XUV}=1.5$ R$_\oplus$ for the purposes of photoevaporation, given the high temperature and low molecular weight of the primordial atmosphere, but cooler and more compact than the more extended atmosphere present immediately after formation.

To compute the total mass loss based on this formula precisely would
require modeling the depth and scale height of the atmosphere over
time. However, a rough estimate can be made by computing the total XUV
radiation absorbed by the planet over its lifetime. For our general results, compute the integrated mass loss over 5 Gyr, which is both the median age of planet host stars in the solar neighborhood \citep{Bonfanti16} and is close to the age of our own solar system. While the stellar XUV flux falls off significantly within a few hundred Myr, we integrate all the way to 5 Gyr for completeness. Note, however, that for the particular case of Earth, planet formation models must account for the apparent loss of the primordial atmosphere at much earlier times.

The integrated XUV flux in our model over 5 Gyr is $5.70\times10^{18}$ erg cm$^{-2}$.
With an efficiency of 10\%, the total mass of hydrogen lost to photoionization from
the prescription in Eq. \ref{eq:ionloss} is $\sim$750 bars. As a
consistency check, in Section \ref{dissociate} we obtain a very similar result with a more precise prescription for the ionizing flux.

XUV irradiation from the central star is not the only potential source
of ionizing radiation in the planet's environment. The galactic
background of FUV radiation, usually taken to be $1.6\times10^{-3}$ erg s$^{-1}$ cm$^{-2}$ (91 -- 200 nm),
is negligible. However, most stars are born in clusters. The mean XUV
flux in the birth cluster is likely to be a few erg s$^{-1}$
cm$^{-2}$, which is subdominant, but fluxes near the center of the
birth cluster may be as much as 100 times greater, comparable to the
flux from the parent star \citep{Fatuzzo08}. As a result, a planet in a solar system
forming in an especially favorable position in the birth cluster may
experience up to twice as much photoevaporation as a planet orbiting
an isolated star. Note also that these results assume the star's X-ray
flux saturates at an age of 100 Myr, in accordance with the \citet{Ribas05} model. If it saturates at an earlier
time, the star's initial XUV flux will be higher, allowing for greater
mass loss in the first 100 Myr of the planet's history.


\subsection{EUV Photo-Dissociation and Evaporation}

\label{dissociate}
We now consider a second mechanism for ultraviolet-induced
photoevaporation. In addition to ionization, longer-wavelength photons
of 91 -- 111 nm (11.2-13.7 eV) are sufficient to photodissociate hydrogen molecules \citep{Draine96}. This dissociation is a
second pathway to input energy into the upper atmosphere and drive
evaporation, analogous to the action of ionization, and suggests that the usual convention of 10\% efficiency of photoevaporation may be underestimated. Both of these
wavelength regimes stand in contrast to the case of longer wavelength photons,
which mostly undergo elastic scattering and do not input energy.

In addition to considering a second ultraviolet heating process, we can obtain a more precise estimate of the XUV flux from the fit of \cite{Ribas05}. For greater accuracy, they broke down their
fit into five wavelength bins, the reddest of which is 92 -- 118 nm. 
Each bin is fit with a function of the form $F = \alpha t_9^\beta$ 
$\rm{erg s^{-1} cm^{-2}}, t > 100 \rm{Myr}$. For this subsection, we
compute the flux of the individual wavelength bins for a more precise
result, which can also be applied to photoionization. We multiply the reddest bin by 0.73 to include only the flux with enough energy to dissociate hydrogen (91 -- 111 nm).

An even more precise model is available from \cite{Claire12}, who
fitted the same parameters to 
\textit{International Ultraviolet Explorer} spectra of solar analogs. However, they found an integrated flux over the 2 -- 118 nm range about 30\% lower
than the \cite{Ribas05} model, so to be generous, we use
the \cite{Ribas05} fit.

For completeness, we note that previous authors have used different 
long wavelength cutoffs for the relevant XUV and EUV bands (cf. \citealt{Ingersoll69,Lee84,WuChen93}).

\begin{table}[htbp]
\caption{Adopted XUV and EUV Flux Prescriptions}
\begin{center}
\begin{tabular}{l|c|r}
\hline
$\lambda$ (nm) & $\alpha$ & $\beta$ \rule{0pt}{2.6ex} \rule[-1.2ex]{0pt}{0pt} \\ [+2pt]
\hline
0.1-2      & 2.40    & -1.92\rule{0pt}{2.6ex} \\ [+2pt]
2-10       & 4.45    & -1.27                                     \\ [+2pt]
10-36     & 13.5    & -1.20                                     \\ [+2pt]
36-92     & 4.56    & -1.00                                     \\ [+2pt]
92-111   & 1.85    & -0.85                                     \\ [+2pt]
\hline
\end{tabular}
\end{center}
\label{fits}
\end{table}

With these fits, we can compute a better estimate of the absorbed XUV
and FUV flux over the Sun's lifetime by integrating over each bin and
adding it to our estimate of a pure blackbody. The integrated fluxes
over each bin are shown in Table \ref{fluxes}. This table also shows the corresponding atmosphere loss to photoevaporation, assuming an efficiency of 10\% at all wavelengths.

\begin{table*}[htbp]
	\caption{Integrated Flux}
	\begin{center}
		\begin{tabular}{l|r|r|r}
			\hline
			$\lambda$ (nm) & Flux over 100 Myr (erg cm$^2$) & Flux over 5 Gyr (erg cm$^2$)  & Atmosphere Loss (bar)  \rule{0pt}{2.6ex} \rule[-1.2ex]{0pt}{0pt} \\ [+2pt]
			\hline
			0.1-2      & $6.30\times10^{17}$ & $1.30\times10^{18}$ &    171\rule{0pt}{2.6ex}           \\ [+2pt]
			2-10       & $2.61\times10^{17}$ & $8.93\times10^{17}$ &     118                                     \\ [+2pt]
			10-36     & $6.75\times10^{17}$ & $2.51\times10^{18}$ &    330                                     \\ [+2pt]
			36-92     & $1.44\times10^{17}$ & $7.07\times10^{17}$ &      93                                     \\ [+2pt]
			92-111   & $4.12\times10^{16}$ & $2.61\times10^{17}$ &      34                                     \\ [+2pt]
			0.1-111  & $1.75\times10^{18}$ & $5.67\times10^{18}$ &     746\rule[-1.2ex]{0pt}{0pt}  \\ [+2pt]
			\hline
		\end{tabular}
	\end{center}
	\label{fluxes}
\end{table*}

With an efficiency factor of 10\%, the combined ionizing and dissociating flux impinging on a young, Earth-like planet
could remove about 750 bars of hydrogen, or about 0.065\% of the mass of our model planet. We
compute the losses due to ionizing flux specifically at 712 bars, 5\% less than our estimate using a single-component model for the stellar flux.

For completeness, we note that using the energy limited approximation, provides an upper limit to the expected evaporation rates. For sufficiently strong radiation fluxes (usually associated with very short-period planets) and hence large mass loss rates, the efficacy of UV heating tends to saturate (e.g., \citealt{Murray-Clay09,OwenAdams}), so that the linear relationship in equation (10) breaks down (the mass loss rate increases more slowly than a linear function of the radiation flux). Although such considerations are beyond the scope of this paper, future work should consider more sophisticated models of this process that go beyond the energy limited regime.

\subsection{Total Mass Loss}

The previous subsections have outlined the various mechanisms
through which mass loss can take place. Jeans escape (Section 3.1)
corresponds to the limiting case where the escaping molecules are
collisionless. Because this process takes place high in the
atmosphere, where the density is low, this mechanism is inefficient.
In the opposite limit where the atmosphere is collisional, the
outflowing material behaves as a fluid. The most restrictive case,
described in Section 3.2 and denoted here as a ``hydrodynamic wind,''
explicitly requires the flow to pass smoothly through its sonic
transition. This hydrodynamic model applies for different heating
mechanisms. We consider the case of heating from both photoionization 
(Section 3.5) and photodissociation (Section 3.6) for the case where
the outflow is energy-limited. We also consider
outflow driven by stellar wind ablation (Section 3.3) and impact
erosion (Section 3.4), where the latter provides a substantial 
contribution.

The results of the models indicate that photodissociation is a non-negligible contribution to mass loss on a young, Earth-like planet formed by pebble accretion, and photoevaporation in general is dominant over all other mechanisms other than impact erosion even before accounting for the potential greater efficiency due to the dissociation contribution to upper atmosphere heating. Because the amount of impact erosion in pebble accretion is uncertain, it is possible that photoevaporation could be dominant. Nonetheless, the total mass loss we compute for all of the processes we study for our model planet is 3.1$\times10^{-3}\,M_\oplus$, or 2.7 kbar, only 15\% of the mass of the initial hydrogen envelope. The contributions of each of these mechanisms to the total are listed in Table \ref{sumtable}.


\begin{table*}[htbp]
	\caption{Mass Loss from All Mechanisms}
	\begin{center}
		\begin{tabular}{l|r|r}
			\hline
			Process & Mass Removed (bars) & Mass Removed ($10^{-6}\,M_\earth$) \rule{0pt}{2.6ex} \rule[-1.2ex]{0pt}{0pt} \\ [+2pt]
			\hline
			Jeans Escape             &     0.6           & 50\rule{0pt}{2.6ex}             \\ [+2pt]
			Stellar Wind Ablation &      20          & 8          \\ [+2pt]
			Impact Erosion\tablenotemark{1}        &  2,300        & 2,000                              \\ [+2pt]
			Photoionization         &   712          & 619                              \\ [+2pt]
			Photodissociation     &     34          & 30                              \\ [+2pt]
			Total                         &   3,067       & 2,700\rule[-1.2ex]{0pt}{0pt}    \\ [+2pt]
			\hline
		\end{tabular}
	\end{center}
	\tablenotetext{1}{Based 0.01 $M_\earth$ of impactor mass.}
	\label{sumtable}
\end{table*}


The difficulty, as noted above, is that planet formation models that seek to explain rapid terrestrial planet formation must strip any primary atmosphere early in the planet's history, and the total mass loss we find for an early Earth analog is not sufficient. It may yet be possible to evaporate the entire primordial atmosphere if the fraction of late planetesimal accretion is greater, closer to 5-10\% of Earth's mass rather than 1\%. Otherwise, the initial gas accretion must be significantly less efficient, leading to a less massive initial atmosphere. The prescription for accretion in \citet{Ginzburg16} suggests that the accreted mass of hydrogen could vary by perhaps a factor of 2. The efficiency of photoevaporation may also be higher, accounting for the overlap between the ionizing and dissociating radiation wavelength ranges, but even an efficiency of 25\% would increase the total mass loss by only 50\%, far from sufficient to evaporate all of the hydrogen and helium. The most plausible scenario in this context would seem to be pebble accretion followed multiple late giant impacts of the Moon-forming type could potentially strip the entire atmosphere.



\section{Discussion and Implications}
\label{discussion}

Given an Earth-like planet formed by pebble accretion that accretes
2\% of its mass in hydrogen and helium from the gas disk, we are
unable to find a clear path to evaporating all of this atmosphere. However, for the purpose of modeling exoplanets, energy-limited photodissociation could be a significant contributor to photoevaporation and should be incorporated into existing models of mass loss. This has significant implications for the location of the evaporation valley in radius-flux space.


Yet it is also true that there are several mechanisms that could reduce the
predicted mass loss. For photoevaporation, a further complication is
that its efficiency may be reduced due to energy loss from line
cooling. Both Ly${\rm \alpha}$ cooling and metal line cooling have
been considered in the case of photoionization
\citep{Murray-Clay09}. When photodissociation is added, molecular
lines must also be included and will further enhance this effect. Of
particular concern is that, for water dissociation, some of the photon
energy will be lost to rotational and vibrational modes of the
hydroxide radical, reducing the mass loss efficiency by $\sim50\%$
over much of the FUV range. However, the relatively low abundance of
water in the primordial atmosphere means this is a negligible effect
compared with the evaporation of hydrogen. Dissociation of molecular
hydrogen does not have this concern because the dissociated atoms have
no molecular lines.

Additionally, magnetic fields are predicted to suppress mass loss in
hot jupiters \citep{adams2011,OwenAdams}. If the surface fields are of
order 1 gauss, then magnetic fields may be sufficient to suppress
outflows from terrestrial planets. This would reduce our expected mass
loss for most processes except for impact erosion, but a full
treatment of this problem is beyond the scope of this paper.

Our photodissociation model of mass loss may also be applied to
close-orbiting exoplanets. Observations indicate that this population
appears to have been significantly sculpted by mass loss
\citep{FP18}. For these highly irradiated planets, photoevaporation is
usually assumed to dominate over impact erosion even in a planetesimal
accretion scenario. This work suggests a possible mechanism for even
greater mass loss than is usually predicted, and further study is needed to determine whether
adding photodissociation to mass loss models improves modeling of the
evaporation valley.

\section{Summary and Conclusions}
\label{summary}

This paper has explored a collection of mass loss mechanisms that can
sculpt the atmospheres of Earth-like planets during their first few
hundred million years of evolution. These calculations are presented
in the context of a fast pebble accretion scenario for terrestrial
planet formation, where the resulting planets can capture a
significant atmosphere before dissipation of the gas disk, but the
results are more broadly applicable.  We review various mechanisms
that could contribute, assess their importance, and discuss their
impact on the resulting composition of the atmosphere. In particular,
we discuss whether the dominant hydrogen and helium envelopes can be
removed, thereby leaving water and other heavy volatile molecules on
the planet. Our primary conclusions can be listed as follows:

\begin{enumerate}
	\item Photodissociation of molecular species can be a significant source of mass loss in the early evolution
          of temperate planetary atmospheres in addition to photoionization.
	\item Impact erosion in a pebble accretion scenario is
          uncertain, but could easily dominate mass
          loss on young terrestrial planets. This process can remove
          $\sim$2,000 bars in our model.
	\item Other sources of mass loss (including Jeans escape,
          traditional forms of a thermal wind, and ablation by stellar
          winds), are unlikely to contribute significantly in the scenario we explore here.
	\item Within the context of our model, where the planet forms
          rapidly in the presence of a gas rich disk, the early Earth
          is expected to develop an atmosphere of $\sim23,000$ bars. However, there is not a clear path to evaporating the bulk of this primordial hydrogen and helium.
	\item The remaining atmosphere would be enriched in water and
          perhaps other volatiles because of the preferential loss of
          hydrogen (and helium) in the outer atmosphere.
\end{enumerate}

More work is needed to understand the timescales (and nature) of
planet formation in the pebble accretion scenario, investigate impact
erosion in a self-consistent way in this context, and further
constrain the distribution of volatile elements in the gas rich disk
during terrestrial planet formation via rapid pebble accretion inside
the ice-line. We further suggest that researchers studying mass loss
in temperate planet atmospheres consider the impact of photodissociation in their models. This initial effort has calculated mass loss rates using a (standard) 
energy limited approximation with a fixed efficiency; future work should generalize this approach, especially to consider efficiency as a function of wavelength given the two processes involved.

\acknowledgements

MRM and ARH are grateful for support from NASA through the JWST NIRCam project though contract number NAS5-02105 (M. Rieke, University of Arizona, PI). 
ARH thanks Ruth Murray-Clay and Eric Lopez for helpful conversations. We also thank Nuria Calvert, and Edwin Bergin for valuable discussions that greatly improved this manuscript. FCA acknowledges support from NASA through the Exoplanets Research Program NNX16AB47G.


\bibliographystyle{apj}
\bibliography{apj-jour,MLrefs}


\appendix
\section{Derivation of Thermal Wind Mass Loss Rate}
\label{derivation}

The rate of thermal wind erosion from a planetary atmosphere can be estimated from a reduced version of the equations of motion where the flow is taken to be isothermal, which is a reasonable approximation for the upper atmosphere. For this case, the solutions for the dimensionless fluid fields can be found analytically, including the required conditions for the flow to pass smoothly through the sonic transition. In order to complete the solution, we must then specify the values for the physical parameters, i.e., the density $\rho_b$ at the exobase (the inner boundary of the flow) and the sound speed $a_s = \sqrt{k_B T/\mu}$.

\subsubsection{Formulation of the Wind/Outflow Problem} 

The equations of motion for this problem include the continuity equation, 
\be
{\partial \rho \over \partial t} + \nabla \cdot \left( 
\rho \uvec \right) = 0 \, ,
\label{continue} 
\ee
the force equation, 
\be
{\partial \uvec \over \partial t} + \uvec\cdot \nabla \uvec = 
- \nabla \Psi - {1\over \rho} \nabla P \, , 
\label{force} 
\ee
and the energy equation 
\be
\rho \left( {\partial \energy \over \partial t} + 
\uvec \cdot \nabla \energy \right) = - P \nabla \cdot \uvec
+ \Gamma - \Lambda \, .
\label{energy} 
\ee

We consider the gravitational potential $\Psi$ to be that of the planet, which is taken to be spherical with mass $M_P$ and radius $R_P$. Since the planet spins, the full potential has an additional contribution from the rotating frame of reference. The order of this correction term is ${\cal O}( \Omega^2 R_P^2/v_{esc}^2)$, which has size $\sim 10^{-3}$ near the planet's surface and can be ignored in this treatment.

In the energy equation (\ref{energy}), $\energy$ is the specific energy of the fluid, $\Gamma$ is the heating rate (per unit volume), and $\Lambda$ is the cooling rate. To start, we consider the gas to be isothermal and replace the energy equation with the simple equation of state
\be 
P = a_s^2 \rho \,.
\ee 

\subsection{Reduced Equations of Motion} 

In this section, we consider steady-state solutions and spherical symmetry. In this regime, the continuity and force equations thus reduce to the forms
\be
{\partial \over \partial r} \left( r^2 \rho v \right) = 0 
\qquad {\rm and} \qquad 
v {\partial v \over \partial r} + {\partial \Psi \over \partial r} 
+ {1\over \rho} {\partial P \over \partial r} = 0 \,.
\label{simple} 
\ee
Next, we assume that the flow is isothermal with constant sound speed $a_s$ and define the following dimensionless quantities,
\be
u \equiv {v \over a_s} , \qquad 
\alpha \equiv {\rho \over \rho_b} , \qquad 
\xi \equiv {r \over R_P} , \qquad {\rm and} \qquad 
\psi \equiv {\Psi \over a_s^2} \, . 
\ee 
Here, $R_P$ is the radius of the planet and $\rho_b$ is the density at the inner boundary $\xi$ = 1. The continuity equation thus takes the form
\be
\alpha {\partial u \over \partial \xi} + 
u {\partial \alpha \over \partial \xi} = 
- {2 \over \xi} \alpha u \, , 
\ee
and the force equation becomes 
\be
u {\partial u \over \partial \xi} + {1 \over \alpha} 
{\partial \alpha \over \partial \xi} = - {\partial \psi \over \partial \xi} 
= - {b \over \xi^2} \,,
\ee
where 
\be
b \equiv {G M_P \over a_s^2 R_P} \,.
\ee
These equations can be integrated immediately to obtain the solutions 
\be
\alpha u \xi^2 = \lambda_{th} \, , 
\label{simplecont} 
\ee
and 
\be
{1 \over 2} u^2 + \log \alpha - {b \over \xi} = \varepsilon \,, 
\label{simpleforce} 
\ee
where the parameters $\lambda_{th}$ and $\varepsilon$ are constant.

\subsection{Sonic Point Conditions} 

In order for the flow to pass smoothly through the sonic point, only particular values of the constant $\lambda_{th}$ are allowed. To quantify this constraint, the boundary conditions at the planetary surface take the form
\be
\xi = 1 \, , \qquad \alpha = 1 \, , \qquad {\rm and} \qquad
u = u_b = \lambda_{th} \,,
\label{innerbc} 
\ee
where the final equality follows from the continuity equation evaluated at the surface. Since $\lambda_{th}$ is determined by the conditions at the sonic point, $u_b$ is specified. The remaining parameter $\varepsilon$ is determined by evaluating the force equation at the inner boundary of the flow, i.e., 
\be
\varepsilon = {1 \over 2} u_b^2 - b 
= {1 \over 2} \lambda_{th}^2 - b \,. 
\label{setenergy}
\ee
The outflow starts with subsonic speeds so that $u_b \ll 1$, whereas typical planet properties imply that $b \sim 5 - 60$. As a result, we can use the approximation $\varepsilon \approx - b$.

For the equations of motion (\ref{simplecont}) and (\ref{simpleforce}), the required matching conditions at the sonic point take the form 
\be
u^2 = 1 \qquad {\rm and} \qquad 
{2 \over \xi} = {b \over \xi^2} 
\quad \Rightarrow \quad \xi_s = {b \over 2} \, . 
\ee

The value of the parameter $\lambda_{th}a$ that allows for smooth flow
through the sonic point is given by
\be
\lambda = {1\over4} b^2 
\exp\left[ {1\over2} \lambda_{th}^2 - b + {3\over2} \right] \,, 
\label{lambda} 
\ee
Equation (\ref{lambda}) provides an implicit solution for the parameter $\lambda_{th}$.  However, the $\lambda_{th}^2$ term on the right hand side of equation (\ref{lambda}) is extremely small (equal to $u_b^2/2 \ll 1$) and can be ignored to leading order; doing so results in an explicit solution for the parameter $\lambda_{th}$, which can be written in the form 
\be
\lambda_{th} \approx { {\rm e}^{3/2} \over 4} b^2 {\rm e}^{-b}\,. 
\ee

\subsection{Estimating the Physical Constants}

The previous subsections specify the solutions for the dimensionless fluid fields, including the necessary conditions for passing smoothly through the sonic point and specification of the dimensionless mass outflow rate ${\dot m}$. In this section, we complete the solution by estimating values for the physical parameters $\rho_b$ and $a_s$ that determine the full mass outflow rate, where 
\be 
{\dot M} = 4\pi r^2 \rho v = 4\pi R^2 \rho_b a_s \lambda_{th} \,. 
\ee 
We first note that the dimensionless potential $b$ can be written 
\be
b = {G M_P \over R_b a_s^2} \approx 62.5 
\left(\frac{M_P}{M_\oplus}\right)\left(\frac{R_b}{R_\oplus}\right)^{-1}\left( {a_s \over 1\, {\rm km} \,\,{\rm s}^{-1}} \right)^{-2} \,
\ee

To estimate the density $\rho_b$ at the base of the flow, we start by assuming that incoming radiation heats the atmosphere down to a layer where the incoming UV photons are optically thick. If the atmosphere is isothermal with scale height $H$, then the number density has the form 
\be
n(z) = n_0 \exp[-z/H] \,, 
\ee
and the optical depth as a function of height $z$ (measured from the planetary surface) takes the form 
\begin{align}
\tau(z) &= \sigmauv \int_z^\infty n(z) dz \\ \nonumber
&= \sigmauv n_0 H \exp[-z/H] = \sigmauv n(z) H \,,
\end{align}
where $\sigmauv\approx 2 \times10^{-18}$ cm$^{-2}$ is the cross section for hydrogen to absorb the incoming UV radiation. Setting $\tau(z)=1$, we thus find the starting estimate for the density 
\be
n_b = {1 \over \sigmauv H} \,.
\ee
For a thin atmosphere, the scale height $H$ is given by 
\be
H = {a_s^2 \over g} = {a_s^2 R^2 \over GM_P} = {R_b \over b} \,. 
\ee
The number density at the base of the flow then becomes 
\be
n_b = {b \over \sigmauv R_b} \approx 
4.9 \times 10^{10} {\rm cm}^{-3} \,
\left( \frac{M_P}{M_\oplus} \right) \left( \frac{R_b}{R_\oplus} \right)^{-2} \left( {a_s \over 1\, {\rm km} \,\,{\rm s}^{-1}} \right)^{-2} \,. 
\ee 

It is thus useful to define a fiducial mass loss rate 
\begin{align}
{\dot M}_0 &= 4\pi R^2 \rho_b a_s \\ \nonumber
&= (4.2\times10^{10} {\rm g}\,\, {\rm s}^{-1})
\left( \frac{M_P}{M_\oplus} \right) \left( {a_s \over 1\, {\rm km} \,\,{\rm s}^{-1}} \right)^{-1} \\ \nonumber
&= (5.3\times10^9 {\rm g}\,\, {\rm s}^{-1}) \left( \frac{M_P}{M_\oplus} \right)^{1/2} \left( \frac{R_b}{R_\oplus} \right)^{1/2} b^{1/2} \,,
\end{align}
where $m$ is the molecular weight for atomic hydrogen (since the dominant heating process we consider is dissociating). Then, taking an optimistic value of $R_b=R_{exo}=12,000$ km. The full mass loss rate can then be written, 
\be
{\dot M} = {\dot M}_0 \lambda_{th} = 
(8.1\times10^9 {\rm g}\,\, {\rm s}^{-1}) 
b^{5/2} {\rm e}^{-b}\,. 
\ee



\end{document}